\documentclass[aps,showpacs]{revtex4}

\usepackage{amsmath}
\usepackage{amssymb}
\usepackage[dvips]{graphicx}
\newcommand{\be}{\begin{equation}}
\newcommand{\ee}{\end{equation}}
\newcommand{\ba}{\begin{eqnarray}}
\newcommand{\ea}{\end{eqnarray}}
\newcommand{\lb}{\label}

\newcommand{\half}{\frac{1}{2}}
\newcommand{\dd}{\partial}
\newcommand{\nn}{\nonumber}
\begin{document}
      
\title{ Circular Orbits in Einstein-Gauss-Bonnet Gravity}
 
\author{Val\'eria M. Rosa\footnote{e-mail: vmrosa@ufv.br}
 }

 \affiliation{
 Departamento de Matem\'atica, Universidade Federal de Vi\c{c}osa, 
 36570-000 Vi\c{c}osa, M.G., Brazil}

\author{ Patricio  S. Letelier\footnote{e-mail: letelier@ime.unicamp.br}
 } 
 
\affiliation{
 Departamento de Matem\'atica Aplicada-IMECC,
 Universidade Estadual de Campinas,
 13081-970 Campinas,  S.P., Brazil}

\begin{abstract}
 The stability under radial and vertical perturbations of circular
orbits associated to particles orbiting a spherically symmetric center
of attraction is study in the context of the n-dimensional: Newtonian
theory of gravitation, Einstein's general relativity, and
Einstein-Gauss-Bonnet theory of gravitation. The presence of a
cosmological constant is also considered.  We find that this constant
as well as the Gauss-Bonnet coupling constant are crucial to have
stability for $n>4$.
\end{abstract}
\maketitle

\section{Introduction}

The physical laws for nonrelativistic phenomena are usually written in
a three dimensional space and for relativistic phenomena in a four
dimensional spacetime. In both cases these laws admit natural
extensions to spaces with arbitrary number of dimensions, these can be
taken as the possible existence of an operative principle (or
principles) which in additions to these laws lead us to determine the
``true dimensions" of the spacetime, the arena where the physical
phenomena occur.

Kant observed that the three-dimensionality of the usual space could
be related with Newton's inverse square law; 
`` The reason for the   three-dimensionality of  the space
is yet unknown. It is likely that the three-dimensionality of space
 results from the
law according to which the forces on the substances act into each other.
The three-dimensionality seems to result from the fact that the substances of
the existing work act into each other in such a way that the strength of
the effect behaves like the inverse of the square of the distances" \cite{kant1}\cite{kant2}.

Since in a first approximation, the galaxies and planetary systems can
be considered as formed by particles moving in circular orbits around
an attraction center, a possible principle to determine the
dimensionality of the space is the existence of stable circular
orbits. The study of the stability of circular orbits moving in
potentials solutions of the n-dimensional Laplace's equations can be
found in \cite{routh}. These studies on stability culminate with the
Ehrenfest work~\cite{ehrenfest} where he shows that in the
Newtonian-Keplerian problem there exists stable limited orbit if and
only if the dimension of the space is three (the potential must be
null at infinity).

In Einstein's gravity Tangherlini~\cite{tangherlini} study the
stability under radial perturbation of orbits associated to test
particles moving around a $n$-dimensional Schwarzschild black hole.
He found stability for $n=4$ only, i.e., in complete concordance with
the Newtonian case. The dimension of the constant time slice, $n-1$,
will be denoted by $d.$

Generalization of the Newton theory that consists in the addition of a
cosmological constant was studied by several authors, see for instance
\cite{bondi} (cosmological context) and \cite{wilkins} (potential
theory context). Clearly this generalization is motivated by the
success of Einstein's gravity with cosmological constant. An example
of model that makes use of a cosmological constant is the $\Lambda$
Cold Dark Matter model ($\Lambda$CDM) \cite{lamcdm} that explains in a
simple way the observed acceleration of the Universe as well as the
cosmic microwave background.

The Hilbert Lagrangian (Ricci scalar times square root of the metric
determinant) is a divergence in two dimensions.  We have that the
Hilbert action is proportional to the Euler characteristic of the two
dimensional manifold (Gauss-Bonnet theorem \cite{manfredo}). In four
dimensions we have that the Bach-Lanczos
Lagrangian~\cite{bach}\cite{lanczos} (a particular combination of
terms quadratic in the Riemann-Christoffel curvature tensor) is a
divergence, so in four dimensions this Lagrangian has no dynamical
significance, like the Hilbert Lagrangian in two dimensions. The
simplest generalization of the Hilbert Lagrangian for $n$-dimensional
spaces ($n>4$) can be obtained adding  the Bach-Lanczos Lagrangian
  to the Hilbert one.  This new theory of gravitation,
Einstein-Gauss-Bonnet (EGB) theory, has some remarkable mathematical
properties, see for instance~\cite{nathalie}.

Another outstanding feature of the Einstein-Gauss-Bonnet theory is
that the Bach-Lanczos action appears as the leading quadratic
correction term of the Hilbert action in the expansion of the
supersymmetric string theory~\cite{deser}.  To be more precise the
Bach-Lanczos action appears exactly in the expansion of the
$E_8\times E_8$ heterotic string model~\cite{duff}.

The aim of this paper is to study the stability under radial
perturbations as well as vertical perturbations of circular orbits
related to test bodies orbiting a spherically symmetric center of
attraction in the above mentioned theories of gravity in an arbitrary
number of dimensions.

 In Section II we study the stability of circular orbits in a
$n$-dimensional Newtonian context with cosmological constant.  In
Section III we consider the stability of circular orbits in the EGB
theory. We divide the study of stability in two. First, in Sect. IV,
we consider the case without cosmological constant. In the next
section (Sect. V) we analyze the case where this constant is not
null. We end summarizing our results in Sect. V.

\section{$d$-dimensional  Newtonian center of attraction 
with cosmological constant}

The Laplace equation with cosmological constant is \be
\nabla^2\Phi+\Lambda=0, \lb{lap} \ee where $\nabla^2$ is the Laplacian
operator in $d$-dimensions, $\Phi$ and $\Lambda$ are the Newtonian
potential and the cosmological constant, respectively.  For spherical
symmetry Eq. (\ref{lap}) reduces to \be
\frac{1}{r^{n-2}}\frac{\dd}{\dd r}(r^{n-2}\frac{\dd}{\dd r}\Phi)
+\Lambda=0 \lb{lap2} \ee that has as solution for $n \geq 4$ , \be
\Phi=- \frac {C}{n-3}\frac{1}{r^{n-3}} - \frac{\Lambda}{2(n-1)} r^2,
\lb{sollap} \ee where $C$ is an integration constant, that for $n=4$
($d=3$) it is proportional to the mass of the center of attraction
($C=GM$, with $G$ the Newton's constant of gravitation).

The Newton equations for an axially symmetric potential can be written
as
 \ba &&\ddot{R}-R\dot{\varphi}^2= -\frac{\dd \Phi}{\dd R}, \lb{n1}\\
&&\frac{d}{dt}(R^2\dot\varphi)=0, \lb{n2}\\ &&\ddot{z^i}= -\frac{\dd
\Phi}{\dd z^i}, \lb{n3} \ea
 where the overdots denote derivation with
respect to $t$, $(R,\varphi, z^1, ... , z^{d-2})$ are $d$-dimensional
cylindrical coordinates, and $\Phi=\Phi(R, z^1, ... ,
z^{d-2})$. Defining the effective potential $\Phi_{eff}=
\Phi+h^2/(2R^2)$, with $h=R^2\dot\varphi$(=constant), we find,
 \be
\ddot{R}= -\frac{\dd \Phi_{eff}}{\dd R}, \; \ddot{z^i}= -\frac{\dd
\Phi_{eff}}{\dd z^i}. \ee
 Now let us study the stability of the
circular orbit $R=R_0, z^i=0$. We shall assume that the potential has
reflection symmetry on the planes $z^i=0$, i.e., $\Phi=\Phi[R,
(z^1)^2, ..., (z^{d-2})^2]$. Expanding the effective potential around
this circular orbit we find,
 \be \Phi_{eff}= \Phi_{eff}(R_0,0, ... ,0) +\half
\frac{\dd^2 \Phi_{eff}(R_0,0, ... ,0)}{\dd R^2}(R-R_0)^2 +
\half\sum_{i=1}^{d-2} \frac{\dd^2 \Phi_{eff}(R_0,0, ... ,0)}{\dd
(z^i)^2}(z^i)^2 + \cdots. \lb{phiexp} \ee 

From Newton's equations
we find the motion for the perturbations $ \rho=R-R_0$ and
$\zeta^i=z^i-0$,
\be \ddot{\rho}+\kappa^2\rho=0,\;
\ddot{\zeta^i}+\nu_{i}^2\zeta^i=0, \lb{freq} \ee 
where
 \ba &&k^2=
\frac{\dd^2 \Phi_{eff}(R_0,0, ... ,0)}{\dd R^2}=\frac{\dd^2
\Phi(R_0,0, ... ,0)}{\dd R^2}+\frac{3}{R}\frac{\dd \Phi(R_0,0,
... ,0)}{\dd R}, \lb{k}\\ &&\nu_{i}^2= \frac{\dd^2 \Phi_{eff}(R_0,0,
... ,0)}{\dd (z^i)^2}=\frac{\dd^2 \Phi(R_0,0, ... ,0)}{\dd (z^i)^2}.
\lb{nu} \ea 
Thus when $\kappa^2>0$ we have that the radial
perturbation remains bounded, i.e., we have linear stability. This
perturbations are usually named as epicyclic perturbations
\cite{binney}. In a similar way when $\nu_{i}^2 >0$ we have that the
``vertical" perturbation $\zeta^i$ also remains bounded. In this case we
say that we have linear stability under perturbations along the
direction $z^i$.  For the spherically symmetric potential
(\ref{sollap}) we find
 \ba &&\kappa^2= (5-n)CR_{0}^{1-n}
-4\Lambda/(n-1), \lb{knew}\\ &&\nu_{i}^2=
\nu^2=CR_{0}^{1-n}-\Lambda/(n-1) \lb{nunew}.  \ea 
To derive the
previous equations we used $r=[R^2+ \sum_{i=1}^{d-2}
(z^i)^2]^{\half}$. Note that because of the spherical symmetry of the
potential we have that all the frequencies for all the vertical
perturbations are equal.

 The relations (\ref{knew}) and (\ref{nunew}) tell us that when
 $\Lambda=0$ we have stability only for $n=4$. For $\Lambda >0$ and
 $n=4$ we have stability whenever $R_{0}^3<3C/(4\Lambda)$. These last
 relation puts an upper limit to the size of the stable galactic
 structures. For $\Lambda >0$ and $n \geq 5$ we have $\kappa^2<0$ and
 the circular orbits are not stable under epicyclic perturbations. Now
 for $\Lambda <0$ we have stable circular orbits for $n=4,5$ and for
 $n \geq 6$ we have stability for $R_0$ limited by the lower bound
 $R_0^*\equiv[(n-5)(n-1)C/(-4\Lambda)]^{1/(n-1)}(<R_{0})$.  In
 other words for certain values of $n$ the presence of a cosmological
 constant can stabilize the circular orbits, sometimes with
 limitations in the size of the orbit.

\section{ Einstein-Gauss-Bonnet theory of gravitation}

The Einstein-Gauss-Bonnet action is,
\be
S=\int d^n x \sqrt{-g}\left [\frac{1}{\kappa^2_n}(R-2\Lambda+\alpha
L_{GB})\right ]+S_{\mbox{matter}},
\label{action}
\ee where $R$ is the $n$-dimensional Ricci scalar ($n>4$), $\kappa_n$
is defined as $\kappa_n \equiv \sqrt{8\pi G_n}$, where $G_n$ is the
$n$-dimensional gravitational constant. We use units such that the
speed of light is one ($c=1$). The Gauss-Bonnet term $L_{GB}$ is given
by \be
L_{GB}=R^2-4R_{\mu\nu}R^{\mu\nu}+R_{\mu\nu\rho\delta}R^{\mu\nu\rho\delta},
\label{GBterm}
\ee and $\alpha$ is a coupling constant. $S_{\mbox{matter}}$ is the
part of the action that describes the matter.  We use the conventions,
$R^a_{bcd}=\partial_c\Gamma^a_{bd}+\cdots$,
$R_{\mu\nu}=R^{\rho}_{\mu\rho\nu}$ and $R=R^{\mu}_{\mu}$.  The action
(\ref{action}) is the simplest one for $n>4$ built with topological
terms. For the general case see \cite{lovelock}.

From (\ref{action}) we find
\be
G_{\mu\nu}-\alpha H_{\mu\nu}-\Lambda g_{\mu\nu}=\kappa^2_nT_{\mu\nu},
\label{gravequation}
\ee
where
\ba
&& G_{\mu\nu}=R_{\mu\nu}-\frac{1}{2}g_{\mu\nu}R, \nn \\ &&
H_{\mu\nu}=2\left[RR_{\mu\nu}-2R_{\mu\alpha}R^{\alpha}_{\nu}-
2R^{\alpha\beta}R_{\mu\alpha\nu\beta}+R_{\mu}^{\alpha\beta\gamma}
R_{\nu\alpha\beta\gamma}\right]-\frac{1}{2}g_{\mu\nu}L_{GB}.
\ea

For vacuum ($T_{\mu\nu}=0$) spherically symmetric solutions to the
EGB equations there is not lost of generality in choosing the metric,
\be ds^2=f(r)dt^2-f(r)^{-1}dr^2-r^2d\Omega^2_{n-2},
\label{metric}
\ee where $d\Omega^2_{n-2}=
d\theta^2_2+\Sigma^{n}_{j=3}(\Pi^{n-1}_{i=2}\sin^2\theta_i)d\theta_j^2\,$  
is the metric of a $(n-2)$-sphere. The case in which this sphere is
changed by a space of negative or null curvature has also been
considered \cite{gleiser}.

From the fact that the  the Einstein tensor, $G^{\mu\nu}$, as well as the Gauss Bonnet tensor, $H^{\mu\nu}$, are divergence free \cite{lovelock} we have that
equation (\ref{gravequation}) tells us that    $T^{\mu\nu}$ is also divergence free. From
this  last fact we conclude that 
 test particles move along geodesics \cite{fock}.
We shall analyze the stability of the circular geodesics $\gamma$
whose parametric equation is, \be t=t(s),\;\; r=r_0,\;\;
\theta_j=\frac{\pi}{2},\;\; j=2,\dots,n-1, \;\; \theta_n=\phi \in
[0,2\pi].
\label{geo}
\ee

The  evolution of a small perturbation, $\xi^{\mu}$, of $\gamma$ is
given by the geodesic deviation equation, \be
\frac{D^2\xi^{\mu}}{ds^2}+R^{\mu}_{\nu\alpha\beta}u^{\nu}\xi^{\alpha}u^{\beta}=0,
\label{geodesv} \ee
where $D/ds=u^\mu\nabla_\mu$, and $u^{\mu}=dx^{\mu}/ds$. Equation
(\ref{geodesv}) can be cast in a simpler, but not manifestly covariant
way, as \be
\ddot{\xi}^{\mu}+2\Gamma^{\mu}_{\alpha\beta}u^{\alpha}\dot{\xi}^{\beta}
+\Gamma^{\mu}_{\alpha\beta,\nu}u^{\alpha}u^{\beta}\xi^{\nu}=0,
\label{syspert}
\ee where the comma denotes ordinary differentiation.  For the curve
(\ref{geo}) the system (\ref{syspert}) reduces to, \ba &&
\ddot{\xi}^0+\frac{f'(r_0)}{f(r_0)}\dot{t}\,\dot{\xi}^1=0, \\ &&
\ddot{\xi}^1+\kappa^2\xi^1=0, \\ && \ddot{\xi}^j+\nu^2\xi^j=0, \\ &&
\ddot{\xi}^n+\frac{2}{r_0}\dot{\phi}\,\dot{\xi}^1=0, \ea where \ba &&
\kappa^2=\frac{f(r_0)(r_0f''(r_0)+3f'(r_0))-2r_0f'(r_0)^2}
{r_0(2f(r_0)-r_0f'(r_0))}, \nn \\ &&
\nu^2=\frac{f'(r_0)}{r_0(2f(r_0)-r_0f'(r_0))}.
\label{frequencies}
\ea

We shall do a study of cases depending on the values of $\alpha$ and
$\Lambda$.  The analysis starts with the stability of circular orbits
for the pure Einsteinian case (no Gauss-Bonnet term). Then we study the
stability of these orbits when a coupling term and/or the cosmological
constant are considered.

\vspace{5mm}

For $\Lambda = 0$ and $\alpha = 0$, the model of spacetime reduces to
the $n$-dimensional Schwarzschild solution. The equation for the
metric function is \be rf'-(n-3)(1-f)=0, \ee which has the general
solution $f(r)=1-C/r^{n-3}$. $C$ is an integration constant that will
taken as positive in order to have the Schwarzschild solution when
$n=4$ ($C=2GM$). In this case the constants $\nu^2$ and $\kappa^2$ can
be cast as, \ba &&
\nu^2=\frac{C(n-3)r_0^{n-3}}{r_0^{n-1}(2r_0^{n-3}-C(n-1))},\nn\\ &&
\kappa^2=\frac{C(n-3)((5-n)r_0^{n-3}-C(n-1))}
{r_0^{n-1}(2r_0^{n-3}-C(n-1))}.\nn \ea

Therefore, we have $\nu^2>0$ if and only if $r_0^{n-3}>C(n-1)$, this imply
that $\kappa^2>0$ if and only if $(5-n)r_0^{n-3}-C(n-1)>0$, and it
only occurs when $n=4$ and $r_0>6GM$. Hence, we have stable circular
orbits only in a 4-dimensional spacetime as it was expected.

\section{ Einstein-Gauss-Bonnet Theory with $\Lambda = 0$.}

Taking $\Lambda = 0$ in (\ref{gravequation}) with $T_{\mu\nu}=0$, the
equation of metric function $f$ is \be rf'-(n-3)(1-f)\left
[-1+\frac{\alpha(n-4)}{r^2}(2rf'-(n-5)(1-f))\right]=0,
\label{fequation} \ee which has the general solution 
\be f_{\pm}(r)=1+\frac{r^2}{2\tilde{\alpha}}\left[1 \pm \sqrt{1+
\frac{4\tilde{\alpha}C} {r^{n-1}}}\right],
\label{fsolution} \ee where $\tilde{\alpha}=(n-3)(n-4)\alpha$ and $C$ 
is an integration constant. The sign of $C$ is chosen analyzing the
behavior of $f$ for $|\alpha|\,\rightarrow \, 0$. Let $f_+$ be the
function $f$ where the sign of square root is positive and let $f_-$
be the other case. When $\alpha$ is small, the asymptotic behavior of
$f_-$ is given by \be f_-\approx 1-\frac{C}{r^{n-3}}.  \ee So,
considering the Gauss-Bonnet term as a perturbation of the
$n$-dimensional Schwarzschild solution, $C$ must be a positive
constant.

When $\alpha$ is small, the asymptotic behavior of $f_+$ is given by
\be
f_+ \approx 1+\frac{r^2}{\tilde{\alpha}}+\frac{C}{r^{n-3}}.
\ee
For $\alpha >0$, it corresponds to Schwarzschild-anti-de Sitter
spacetime with negative gravitational mass, with the standard energy
definition in background anti-de Sitter \cite{abott}.

Then, there are two families of solutions corresponding to the sign in
front of the square root that appears in (\ref{fsolution}). Following
the definition given in \cite{nozawa}, the family with minus (plus)
sign will be called general relativity (GR) branch (non-GR branch)
solution.  The spacetime structure of these solutions of the EGB
theory is studies in \cite{torii}.

\subsection{ The GR branch solution.}

By chosing the solution  $f=f_-$ of (\ref{fequation}) from (\ref{frequencies}) 
we find,

\ba
&& \nu^2=\frac{r_0^n(1-\sqrt{\beta})+Cr_0\tilde{\alpha}(5-
n)}{\tilde{\alpha}(r_0^3C(n-1)-2r_0^n\sqrt{\beta})},\nn\\
&& \kappa^2=\frac{A_1+A_2+A_3}{-2r_0^n\beta\sqrt{\beta}\tilde{\alpha}(
-2r_0^n\sqrt{\beta}+r_0^3C(n-1))},
\label{frequencies1}
\ea
where $\beta=1+4\tilde{\alpha}Cr_0^{1-n}$ and 

\ba
&& A_1=8r_0^{2n}(1-\sqrt{\beta})+r_0^{n+3}C(n^2-1)(1
-\sqrt{\beta}), \nn \\
&& A_2=2r_0^{n+1}[32-\sqrt{\beta}(n^2-8n+39)]+2r_0^4C^2\tilde{\alpha}(n
-1)[n+7-\sqrt{\beta}(n+3)], \nn \\
&& A_3=4r_0^2C^2\tilde{\alpha}^2[32-2\sqrt{\beta}\tilde{\alpha}C(n-5)(n
-9)]-8r_0^{5-n}C\tilde{\alpha}(n-5)(n-1).
\label{den2K0}
\ea

\begin{figure} 
 \includegraphics[scale=.3]{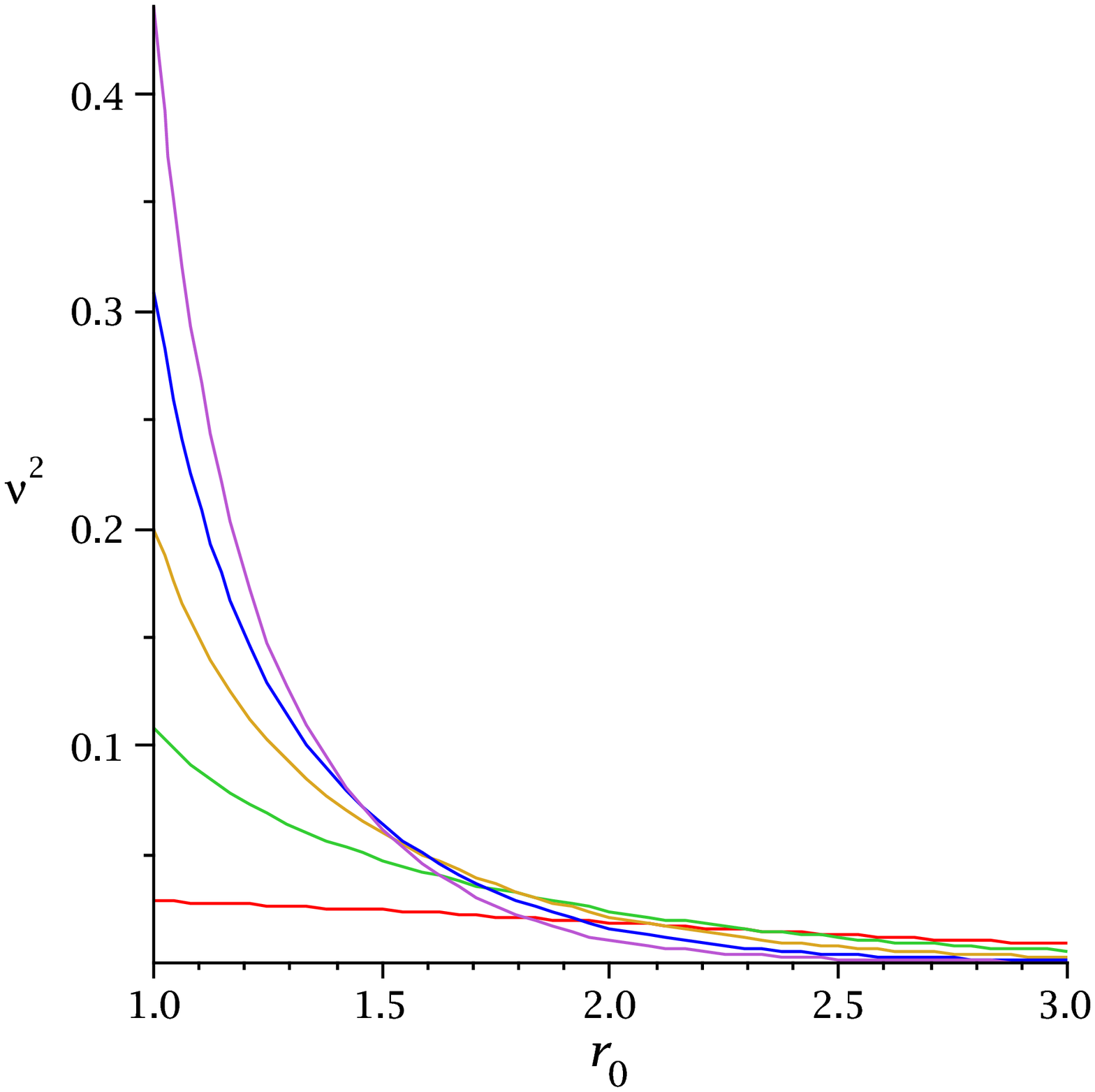} 
 \includegraphics[scale=.3]{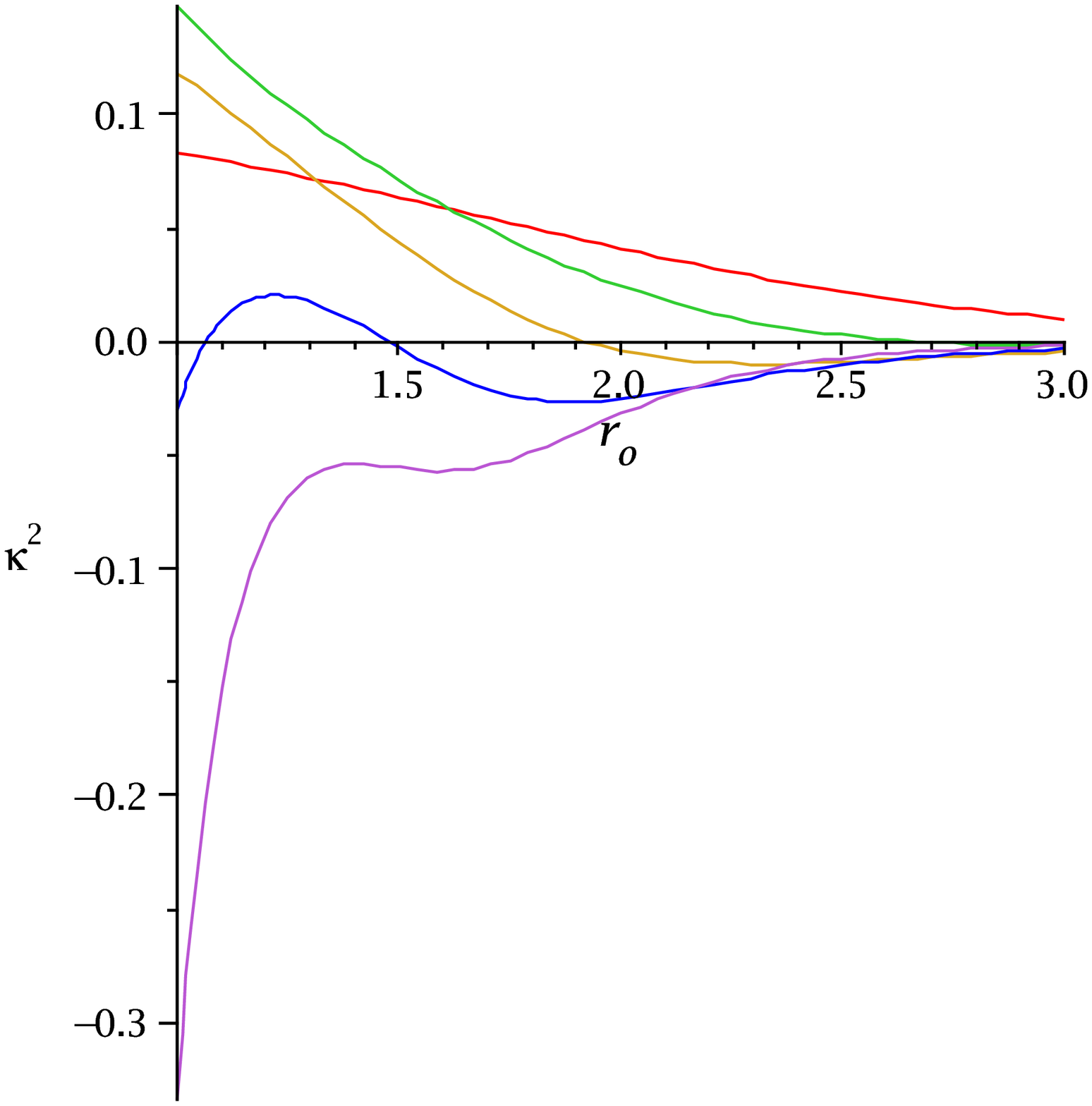}
 \includegraphics[scale=.3]{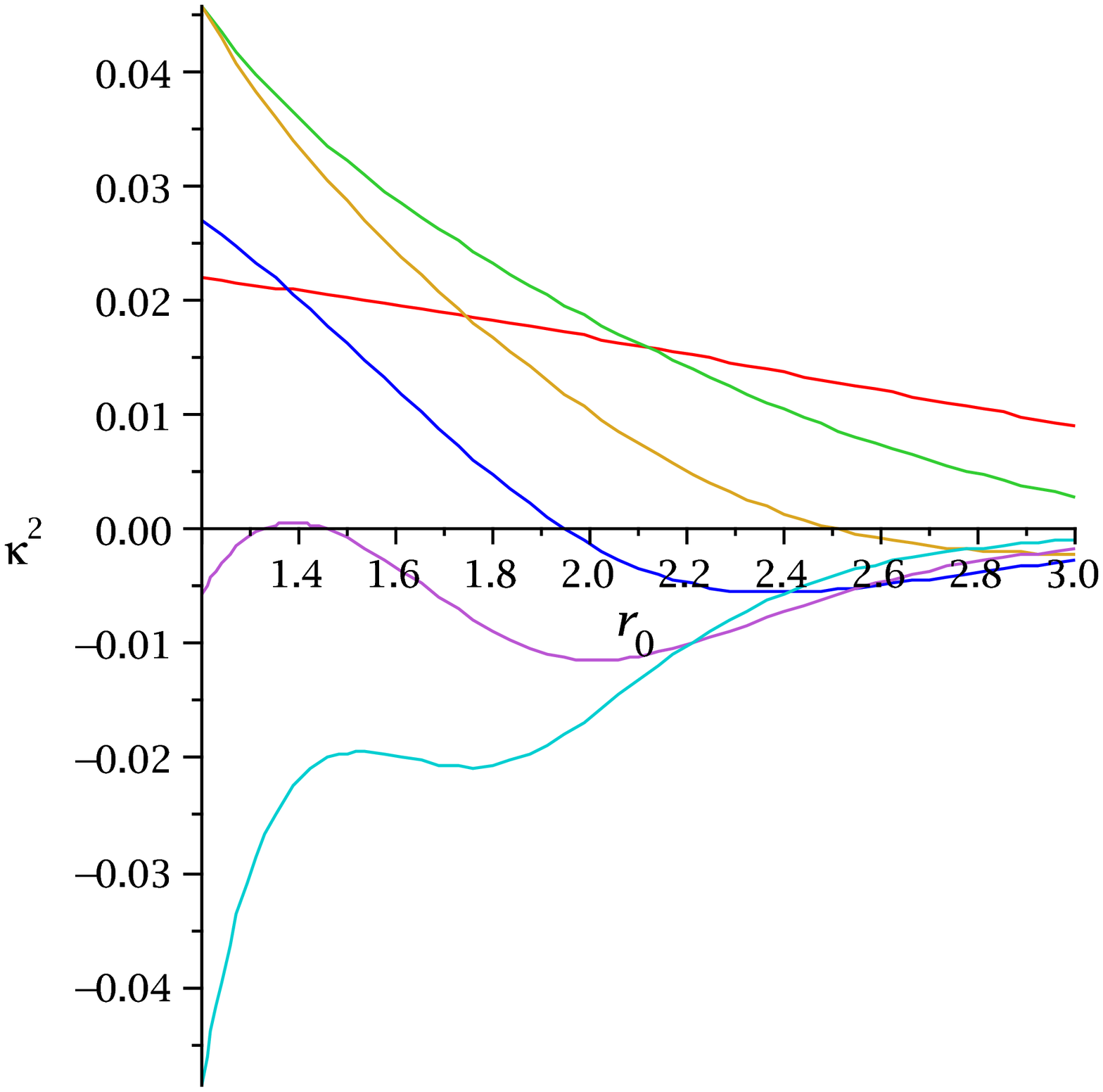}
 \caption{GR branch, $\Lambda = 0$ case. Epicyclic and vertical frequencies   for several values of $n$ and $C=1,$  $1 \le r_0\le 3$.  For the first two figures $\tilde{\alpha}=20$  and for the third $\tilde{\alpha}=80$. In the first graphic $n$ take the values  $n=9$ (top curve), $ 8, 7, 6$ and $5$ (bottom curve).
In the second, counting along the line $r_0=2$, $n$ takes the values $5$ (top curve), $6, 7, 8$ and $9$ (bottom curve). And for the third graphic, counting along the same line, $n= 6$ (top line), $5, 7, 8, 9, 10$ (bottom line). We see that the region of stability shrinks and disappears for large  $n$. }
\label{fig1to3}
 \end{figure}

The behavior of the functions $\nu^2$ and $\kappa^2$ are not simple,
but it is possible to find $n$, $\tilde{\alpha}$, $C$ and $r_0$ such
that $\nu^2$ and $\kappa^2$ are both positive, i.e., there are stable
circular orbits. For example, we have stable circular orbits when
$n=5$, $\tilde{\alpha}=2$, $C=1$ and $0.2 \le r_0 \le 2.0$; $n=6$,
$\tilde{\alpha}=20$, $C=1$ and $0.2 \le r_0 \le 2.0$; $n=7$,
$\tilde{\alpha}=20$, $C=1$ and $0.7 \le r_0 \le 1.9$; $n=8$,
$\tilde{\alpha}=20$, $C=1$ and $1.07 \le r_0 \le 1.47$; and $n=9$,
$\tilde{\alpha}=80$, $C=1$ and $1.33 \le r_0 \le 1.46$. In all these
examples we have $f_-(r_0)>0$ in the specified region. Hence, the
Gauss-Bonnet term can stabilize circular orbits in some higher
dimensions spacetimes.

In   Fig. \ref{fig1to3}  we show 
the epicyclic and vertical frequencies   for several values of $n$ and $C=1,$  $1 \le r_0\le 3$.  For the first two figures $\tilde{\alpha}=20$  and for the third $\tilde{\alpha}=80$. In the first graphic $n$ take the values  $n=9$ (top curve), $ 8, 7, 6$ and $5$ (bottom curve). 
In the second graphic, counting along the line $r_0=2$, $n$ takes the values $5$ (top curve), $6, 7, 8$ and $9$ (bottom curve). There is no stable circular orbit when $n=9$. And for the third graphic, counting along the same line, $n= 6$ (top line), $5, 7, 8, 9, 10$ (bottom line). 
In this case here is no stable circular orbit when $n=10$, but for $n=9$ we have stability.

The vertical frequency, $\nu^2$, is always
  positive no matter the value of $n$ but the range of $r_0$ where
  $\kappa^2 $ is positive decreases when the dimension $n$ increases. 

For $n\ge10$ and $\tilde{\alpha}>0,$ and also for $n\ge 5$ and
  $\tilde{\alpha}<0$, we calculate the functions $\nu^2$ and
$\kappa^2$ into a grid of points with $0.1 \le r_0 \le 2$ (step
$0.1$), $5 \le n \le 20$ and $0.01 \le \tilde{\alpha} \le 100$ (step
$0.01$ into $[0.01,\,0.1]$, step $0.1$ into $[0.1,\,1]$, step $1$ into
$[1,\,100]$) and $C=1$. In the case $\alpha <0$ we work with $-100 \le
\tilde{\alpha} \le -0.01$, with the same steps for the corresponding
intervals. In all these numerical examples we obtain $\nu^2\kappa^2<0$
and, therefore, no circular orbit is stable.

\subsection{ The non-GR branch solution.}

For completeness we study the case when $f=f_+$ is the solution of
(\ref{fequation}). In this case the constants $\nu^2$ and $\kappa^2$
are given as in (\ref{frequencies1}) and (\ref{den2K0}) replacing
$-\sqrt{\beta}$ by $\sqrt{\beta}$.

Also in this case, the stability depends on the sign of $\alpha$. When
$\alpha < 0$, in the expression for $\nu^2=N_\nu/D_\nu$, we have a
numerator, $N_{\nu}=r_{0}^n
[(1+\sqrt{\beta})r_0^n-(n-5)C\tilde{\alpha}r_0]$, non negative and a
denominator, $D_{\nu}=\tilde{\alpha}(r_0^3C(n-1)+2r_0^n\sqrt{\beta})$,
always negative. So, $\nu^2$ is not positive, hence there are no
stable circular orbits.

\vspace{3mm}

\begin{figure} 
 \includegraphics[scale=.3]{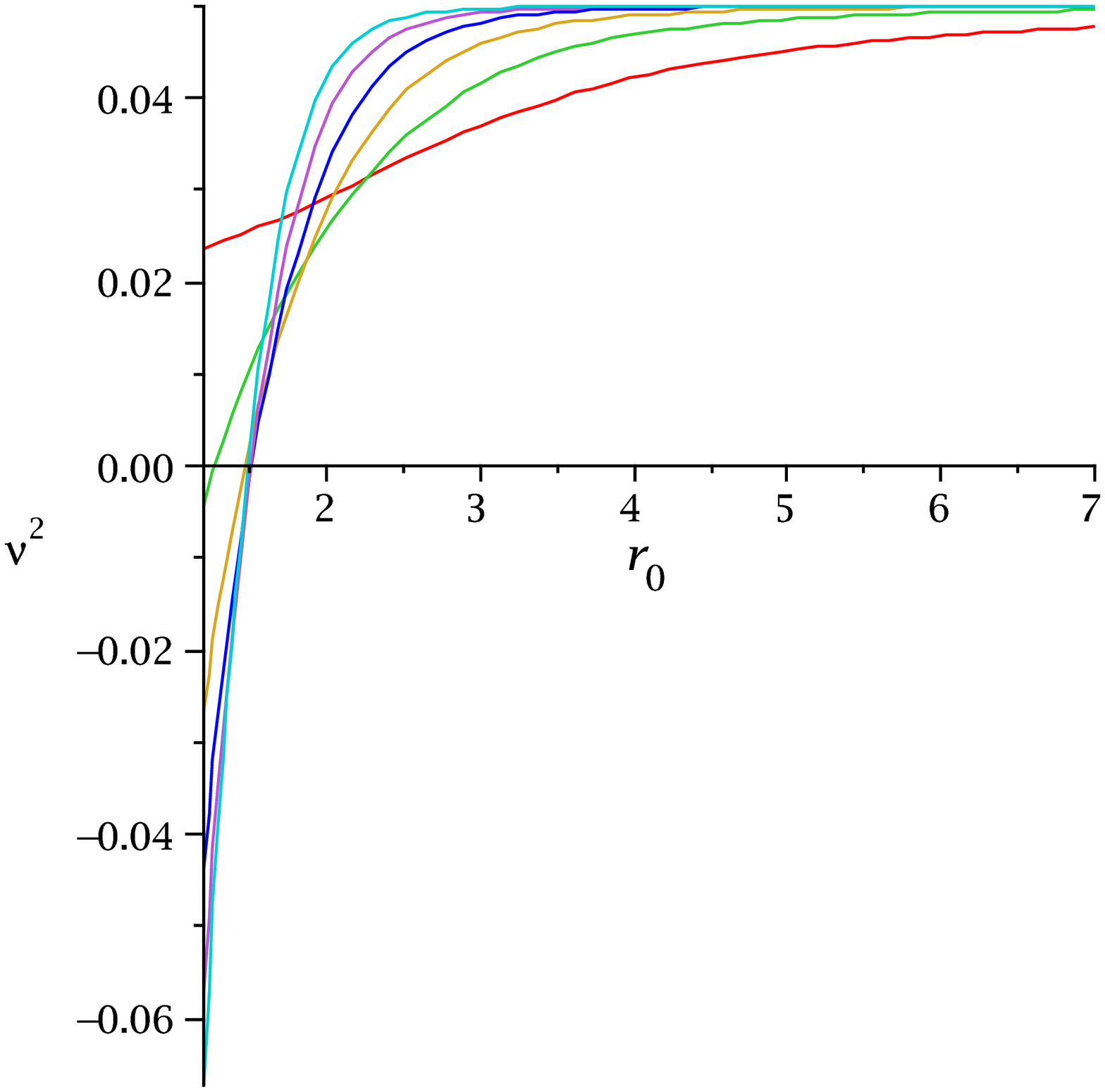} 
 \includegraphics[scale=.3]{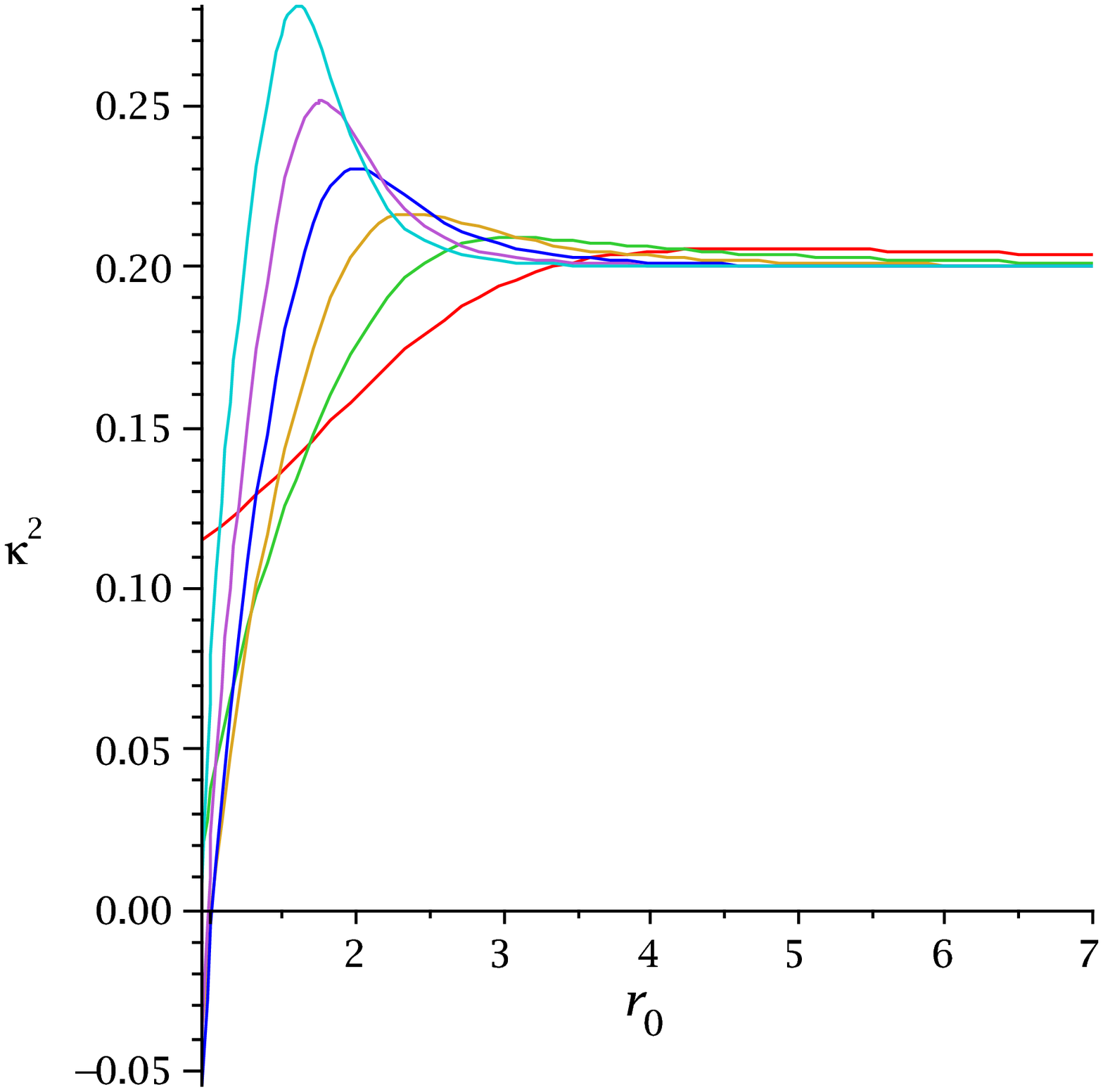}
 \caption{\footnotesize Non-GR branch, $\Lambda = 0$ case. Graphics
of the functions $\nu^2$ and $\kappa^2$ for
   $\tilde{\alpha}=20$, $C=1$ and $1 \le r_0 \le 7$ and    $n=10$ (top line), $\,9,\,8,\,7,\,6,$ $\,5$ (bottom line). The have a large region of stability.}
 \label{fig4to5}
 \end{figure}

The parameters $n(\ge 5)$ and $C$ are positive. When $\alpha>0$ the
denominators of $\nu^2$ and $\kappa^2$ are always positive
($D_{\kappa}=2\beta\sqrt{\beta}r_0^nD_{\nu}$). The same happens with
the functions $f_+$, $\beta$, and with the coefficients of $r_0$ in
$\bar{A_1}$ and $\bar{A_2}$ (Note that $\bar{A_i}$ is the function
$A_i$ after replacing $-\sqrt{\beta}$ by $\sqrt{\beta}$,
$i=1,\,2,\,3$). These positive coefficients assure that, for $r_0$
sufficiently large, the numerator of $\nu^2$,
$N_{\nu}=[(1+\sqrt{\beta})r_0^n-(n-5)C\tilde{\alpha}r_0]$, and the
numerator of $\kappa^2$, $N_{\kappa}=[\bar{A_1}+\bar{A_2}+\bar{A_3}]$,
are positive. Hence, we conclude that it is possible to find stable
circular orbits for $n \ge 5$ and $\alpha > 0$.

 Fig. \ref{fig4to5} shows the functions $\nu^2$  and
  $\kappa^2$ , for $C=1$, $\tilde{\alpha}=20$, $1 \le r \le 7$    and $n=10$ (top line), $\,9,\,8,\,7,\,6,$ $\,5$ (bottom line). We have stability 
  for $r_0 > r_n$, where $r_n$ depends on the dimension $n$.

\section{ Einstein-Gauss-Bonnet Theory with $\Lambda \ne 0$.}

Taking $\alpha=0$ and $\Lambda \ne 0$ in (\ref{gravequation}) with
$T_{\mu\nu}=0$, the metric function obeys the equation, 
\be
rf'-(n-3)(1-f)+\frac{2\Lambda r^2}{n-2}=0, \ee 
which has the general
solution, \be f(r)=1-\frac{C}{r^{n-3}}-\frac{2\Lambda
r^2}{(n-1)(n-2)}, \ee 
the integration constant $C$ will be considered positive as before. 
The frequencies  (\ref{frequencies}) in this case are
$\nu^2=N_1/D_1$ and $\kappa^2=N_2/D_2$, where \ba && N_1=Cr_0^2
(n-1)(n-2)(n-3)^2-4\Lambda r_0^{n+1};\;\;\;
D_1=r_0^4(n-1)(n-2)(2r_0^{n-3}-C(n-1)), \nn \\ && N_2=Cr_0^2
(n-2)(n-3)^2((5-n)r_0^{n-3}-C(n-1))-2\Lambda
r_0^{n+1}(8r_0^{n-3}-C(n^2-1)), \nn \\ &&
\hspace{7cm}D_2=r_0^{n+1}(2r_0^{n-3}-C(n-1))(n-1)(n-2).\ea

\vspace{3mm}

When $\Lambda<0$, the numerator $N_1$ of $\nu^2$ is positive for all
$r_0$, and its denominator $D_1$ is positive if and only if
$r_0^{n-3}>C(n-1)/2$. Note that $D_2$ change of sign with $D_1$. The
function $f$ is positive when $r_0^{n-3}>C$, and this condition is
satisfied when $D_1>0$. So, it is necessary to find the values of
$r_0$ for which the numerator $N_2$ of $\kappa^2$ is positive. For
better visualization, we write $\Lambda=-\lambda$ in $N_2$ \be
N_2=16\lambda
r_0^{2n-2}-2C\,\lambda(n^2-1)r_0^{n+1}-C(n-2)(n-3)(n-5)r_0^{n-1}
-C^2(n-2)(n-3)(n-1)r_0^2.\nn \ee

 Analyzing the variation of the signs of the polynomial $N_2$, we can
see that there is at least one positive root for $N_2$. Adding the
fact that for large $r_0$, $N_2$ is positive, we can conclude that it
is possible to find a value $\bar{r}$ such that $\nu^2>0$ and
$\kappa^2>0$ for $r_0>\bar{r}$. Therefore, when $\Lambda < 0$ and for
$n \ge 4 $ there exist stable circular orbits.

\vspace{3mm}

When $\Lambda>0$, in order to have $f>0$ it is necessary that, at
least, \be 2r^{n-3}-C(n-1)>0. \label{D1} \ee Under this condition the
denominators $D_1$ and $D_2$ are positive. Now it is necessary that
the numerators $N_1$ and $N_2$ be also positive. The polynomial $N_1$
is positive when \be Cr_0^2 (n-1)(n-2)(n-3)^2-4\Lambda r_0^{n+1}>0.
\label{N1}
\ee In four dimensions we have $N_1 > 0$ and $N_2 > 0$ when we
take, for example, $C=0.1$, $\Lambda=0.05$ and $0.32 \le r_0 \le
0.55$. But when $n \ge 5$ the numerators, $N_1$ and $N_2$, can not be
positive at same time. We can see this by doing an analysis of the
polynomial $N_2$. Using (\ref{D1}) and (\ref{N1}) we obtain \ba &&
N_2<-2\Lambda r_0^{n+1}(4C(n-1)-C(n^2-1))-Cr_0^2
(n-2)(n-3)^2((5-n)C(n-1)/2-C(n-1))= \nn \\ && = C(n-1)(n-3)[4\Lambda
  r_0^{n+1}-Cr_0^2 (n-1)(n-2)(n-3)^2]/2<0.  \ea So, for $n \ge 5$ and
$\Lambda>0$ there is no stable circular orbit. It is interesting to
note that the addition of a negative cosmological constant stabilize,
former not stable, circular orbits for any $n \ge 4$.

\vspace{3mm}

In the following we analyze the stability of the circular orbits when
both $\Lambda$ and $\alpha$ are not null. In this case the equation of
metric function is written as
\be
rf'-(n-3)(1-f) +\frac{2\Lambda
r^2}{n-2}+\frac{\alpha(n-4)(n-3)(1-f)}{r^2}(2rf'-(n-5)(1-f))=0,
\label{fequation2}
\ee
which has the general solution 
\be
f(r)\pm=1+\frac{r^2}{2\tilde{\alpha}}\left[1 \pm
\sqrt{1+\frac{4\tilde{\alpha}}{(n-1)(n-2)}\left
(2\Lambda+\frac{C}{r^{n-1}}\right)}\right],
\label{fsolution2}
\ee where $\tilde{\alpha}=(n-3)(n-4)\alpha$ and $C$ is a integration
constant. As before, let $f_+$ be the function $f$ when the sign of
square root is positive and let $f_-$ be the other case. In the limit
$|\alpha|\,\rightarrow \, 0$ the function $f_-$ reduce to \be
f_-\approx 1-\frac{2\Lambda
  r^2}{(n-1)(n-2)}-\frac{C}{(n-1)(n-2)r^{n-3}}.  \ee By considering
the Gauss-Bonnet term as a perturbation of the Schwarzschild-dS
(Schwarzschild-adS) spacetime with $\Lambda >0$ ($\Lambda <0$), we
conclude that $C$ must be positive.

When $\alpha$ is small, the asymptotic behavior of $f_+$ is given by
\be
f_+ \approx 1+\frac{r^2}{\tilde{\alpha}}+\frac{2\Lambda
r^2}{(n-1)(n-2)}+\frac{C}{(n-1)(n-2)r^{n-3}}.
\ee

As we can see, here again, there are two families of solutions
corresponding to the sign in front of the square root in
(\ref{fsolution2}). We define two branches following the same
criterion, the GR branch and the non-GR branch.

\subsection{ The GR branch solution ($\Lambda \ne 0$)}

The functions $\nu^2$ and $\kappa^2$ in this case can be written as
(\ref{frequencies}) \ba &&
\nu^2=\frac{r_0^n[(n-1)(n-2)(1-\sqrt{\beta_*})+8\Lambda\tilde{\alpha}
-\tilde{\alpha}Cr_0^{1-n}(n-5)]}{\tilde{\alpha}(n-1)
(-2r_0^n\sqrt{\beta_*}(n-2)+r_0^3C)},\\ &&
\kappa^2=\frac{B_4r_0^{n+3}+B_3r_0^{n+1}+B_2r_0^n+B_1r_0^4+
B_0r_0^2}{2\tilde{\alpha}(n-1)(-2r_0^n\sqrt{\beta_*}
(n-2)+r_0^3C)r_0^n[(n-1)(n-2)+4\tilde{\alpha}(2\Lambda +Cr_0^{1-n})]}
\lb{frequencies2} \ea where \be
\beta_*=1+\frac{4\tilde{\alpha}}{(n-1)(n-2)}\left
(2\Lambda+\frac{C}{r^{n-1}}\right), \ee and \ba && B_4=
C(8\Lambda\tilde{\alpha}+(n-1)(n-2))(n^2-1)[-\sqrt{\beta_*}+1], \nn \\
&& B_3=
C[-32\tilde{\alpha}\sqrt{\beta_*}(n-1)(n-2)+2\tilde{\alpha}(n^2
-8n+39)(8\tilde{\alpha}\Lambda+(n-1)(n-2))], \nn \\ && B_2=
8\big[[(-\sqrt{\beta_*}+1)(n-1)(n-2)+8\tilde{\alpha}\Lambda][(n
-1)(n-2)+8\tilde{\alpha}\Lambda]\big], \nn \\ && B_1=
2C^2\tilde{\alpha}(n-1)[-\sqrt{\beta_*}(5-n)+n+3], \nn \\ && B_0=
4\tilde{\alpha}^2C^2[(n-5)(n-9)].  \ea

First, when $\Lambda < 0$ it is possible to find stable circular
orbits. For example, if $\tilde{\alpha}=\pm 1$, $\Lambda=-1$, $C=1$ then
$\gamma$ is stable when $r_0 \ge 1.2$ ($n \ge 5$). Note that $f_-=0$
if and only if \be
p(r)=2(-\Lambda)r^{n+4}+(n-1)(n-2)r^{n+2}+(n-1)(n-2)\tilde
{\alpha}r^n-Cr^5=0. \ee Analyzing the variation of signs in the
coefficients of $p$, we can see that, if $\Lambda<0$ then the number
of variations is one and there is only one positive value $\bar{r}$
such that $p(\bar{r})=0$. And for $r_0>\bar{r}$ we have $f(r_0)>0$. If
$\Lambda>0$, the existence of a root of $p(r)=0$ depends on the
relation between $\tilde{\alpha}$, $\Lambda$ and C. In the cases
$\tilde\alpha=\pm1$, $\Lambda=-1$, $C=1$, the polynomial
$p(r)=2r^{n+4}+(n-1)(n-2)(r^2-1)r^n-r^5$ has only one positive
root. Hence, we have a horizon (at $r_H<1$) with the zone of stability
outside this horizon.

\begin{figure} 
 \includegraphics[scale=.3]{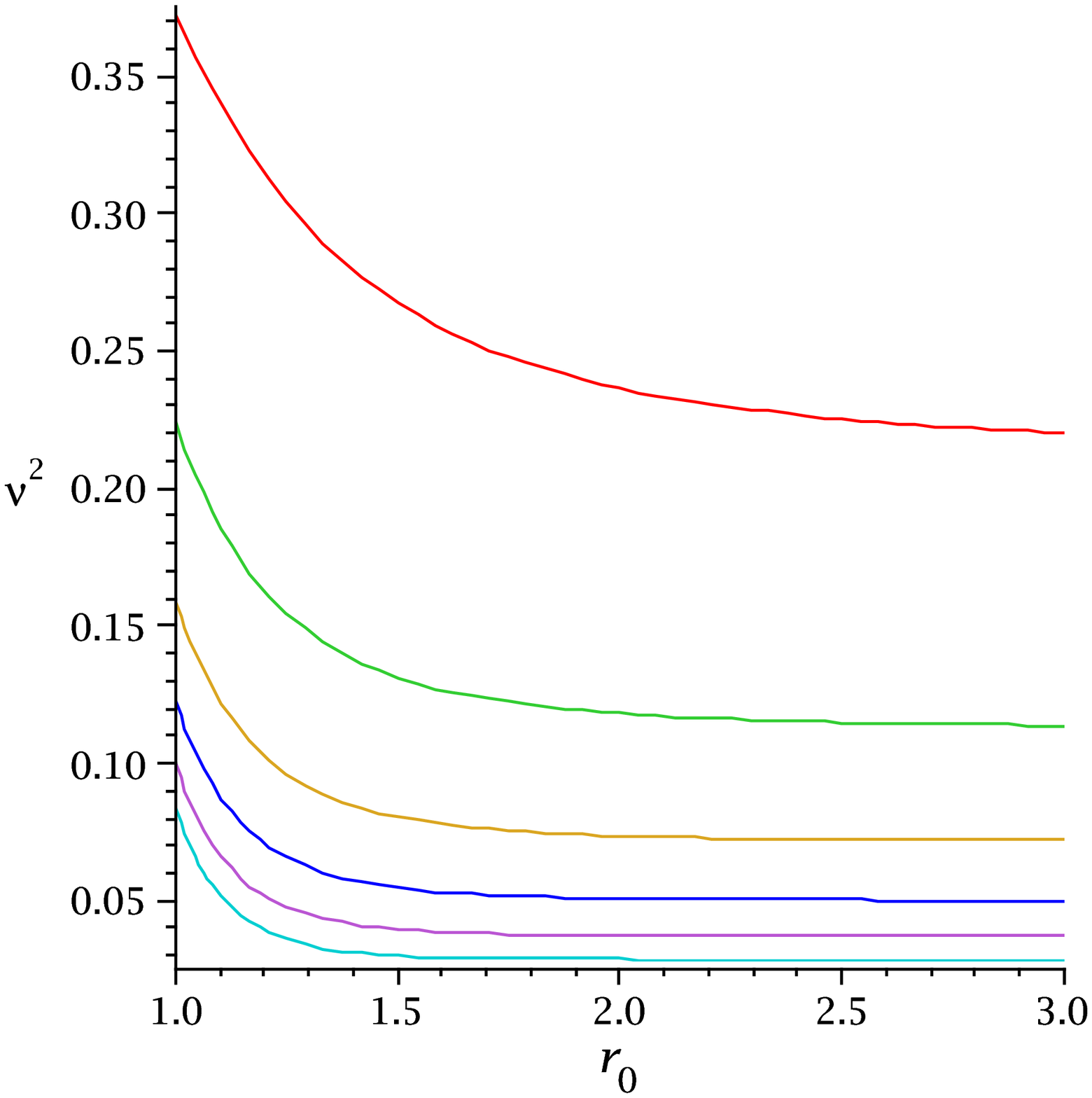} 
 \includegraphics[scale=.3]{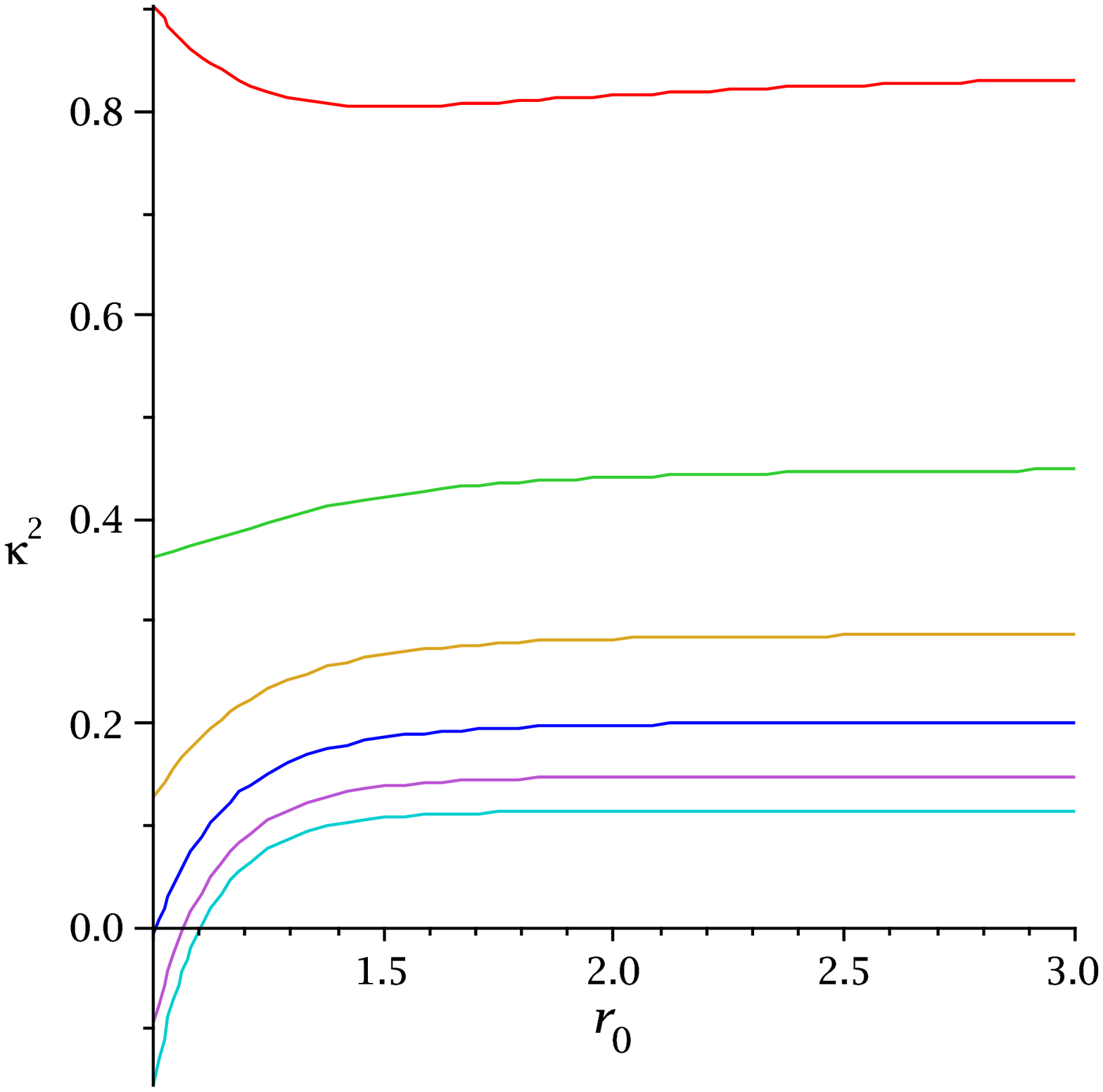}
 \caption{ GR branch, $\Lambda \ne 0$ case.  The functions $\nu^2$  and
   $\kappa^2$  for $\Lambda=-1$, $C=1$, $\tilde{\alpha}=1$
   and $1 \le r_0 \le 3$, and  $n=5$(top curve), $6, 7,
   8, 9, 10$ (bottom curve). For large $n$ we see a small region of instability.}
 \label{fig6to7}
 \end{figure}

Figure \ref{fig6to7}  shows  the behavior of
  the functions $\nu^2$ and $\kappa^2$  for
  $\Lambda=-1$, $C=1$, $\tilde{\alpha}= 1$ and $1 \le r_0 \le 3$, and
    $n=5$(top curve), $6, 7,  8, 9, 10$ (bottom curve). We see a small region of instability for large $n$.  Similar behavior is found for the same parameters,
but $ \tilde\alpha=-1$.

The cases where $\Lambda >0$ was numerically analyzed.  For example,
we have stable circular orbits when $n=5$, $\tilde\alpha = C = \Lambda = 1$
and $0.1 \le r_0 \le 0.5$; $n=6$, $\tilde\alpha = C = \Lambda = 1$ and $0.2
\le r_0 \le 0.55$; $n=7$, $\tilde\alpha = 2$, $C = \Lambda = 1$ and $0.3 \le
r_0 \le 0.55$; $n=8$, $\tilde\alpha = 9$, $C= \Lambda = 1$ and $0.45 \le r_0
\le 0.55$, and $n=9$, $\tilde{\alpha}=100$, $C=1$, $\Lambda=0.05$ and
$0.62 \le r_0 \le 0.83$. In all these examples we have $f_-(r_0)>0$
into the fixed intervals. 

The other cases with $\Lambda >0$ ($n \ge 10$ and $\alpha>0$ or $n \ge
5$ and $\alpha<0$) we calculate the functions $\nu^2$ and $\kappa^2$
into a grid of points $(r_0,\Lambda,\tilde{\alpha},n)$ and at each
point of the grid these function are not positive at same time. The grid
is built with $0.1 \le r_0 \le 2$ (step $0.1$), $0 \le \Lambda \le 2$
(step $0.5$), $5 \le n \le 20$ and $0.01 \le \tilde{\alpha} \le 100$
(step $0.01$ into $[0.01,\,0.1]$, step $0.1$ into $[0.1,\,1]$, step
$1$ into $[1,\,100]$) and $C=1$. In the case $\alpha <0$ we work
with $-100 \le \tilde{\alpha} \le -0.01$, with the same steps for the
corresponding intervals. 

In these cases ($\Lambda > 0$), each value of $n$ has a different set
of parameters $(r_0,\,\tilde{\alpha},\,C)$ where both $\nu^2$ and
$\kappa^2$ are positive, and this differences turns quit impossible to
repeat the graphical analysis as in the $\Lambda <0$ cases.

\subsection{ The non-GR branch solution ($\Lambda \ne 0$)}

Again, for completeness we describe the case where $f_+$ is the
solution of (\ref{fequation2}). In this case, the functions $\nu^2$
and $\kappa^2$ (\ref{frequencies}) are given as (\ref{frequencies2})
replacing $-\sqrt{\beta_*}$ by $\sqrt{\beta_*}$.

The condition to existence of $f$ ($\beta_*>0$) gives us
$[(n-1)(n-2)+8\Lambda\tilde{\alpha}+4\tilde{\alpha}Cr_0^{1-n}]>0$. For
$\alpha <0$ the numerator of $\nu^2$,
$N_{\nu}=r_0^n[(n-1)(n-2)(1+\sqrt{\beta_*})+8\Lambda\tilde{\alpha}
-\tilde{\alpha}Cr_0^{1-n}(n-5)]$, is always positive but the
denominator, $D_{\nu}=\tilde{\alpha}(n-1)
(+2r_0^n\sqrt{\beta_*}(n-2)+r_0^3C)$, is negative, so $\nu^2<0$, and
in this case no circular orbit is stable.

For $\alpha > 0$ we first analyze the case $\Lambda > 0$. In this case
all parameters are positive and the denominators of $\nu^2$ and
$\kappa^2$ are always positive. The same happens with $f_+$,
$\beta_*$, $\tilde{B_4}$, $\tilde{B_3}$, $\tilde{B_2}$ and the
expression $[(n-1)(n-2)(1+\sqrt{\beta_*})+8\Lambda\tilde{\alpha}]$
(Note that $\tilde{B_i}$ is the coefficient $B_i$ after replacing
$-\sqrt{\beta_*}$ by $\sqrt{\beta_*}$, $i=2,\,3,\,4$). These positive
 coefficients
assure that, for $r_0$ sufficiently large, the numerator of $\nu^2$
and $\kappa^2$ are positive. Hence, we conclude that there are stable
circular orbits for $n \ge 5$, $\alpha > 0$ and $\Lambda > 0$.

If $\alpha >0$ and $\Lambda <0$, it is necessary to set conditions to
assure $\beta_*>0$. It happens when
$(n-1)(n-2)+8\Lambda\tilde{\alpha}>0$, what is true when
$2|\Lambda|\tilde{\alpha} < 3$. In this case, from the
expressions for $\nu^2$ and $\kappa^2$, we can see that the same
argument used in the case $\Lambda > 0$ is valid. Hence, we
conclude that there are stable circular orbits for $n \ge 5$,
 $\alpha > 0$ and $\Lambda < 0$.

\begin{figure} 
 \includegraphics[scale=.3]{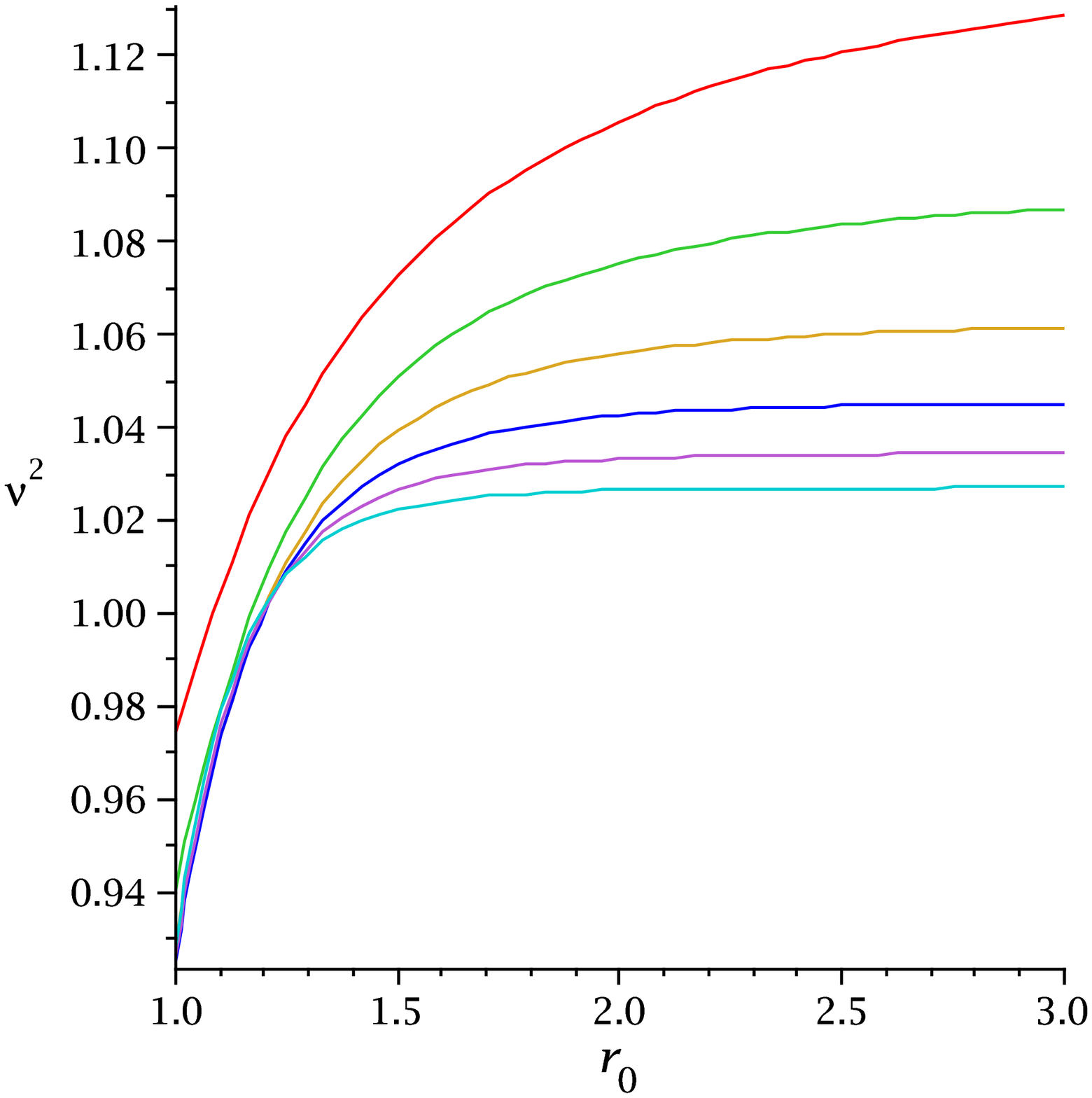} 
 \includegraphics[scale=.3]{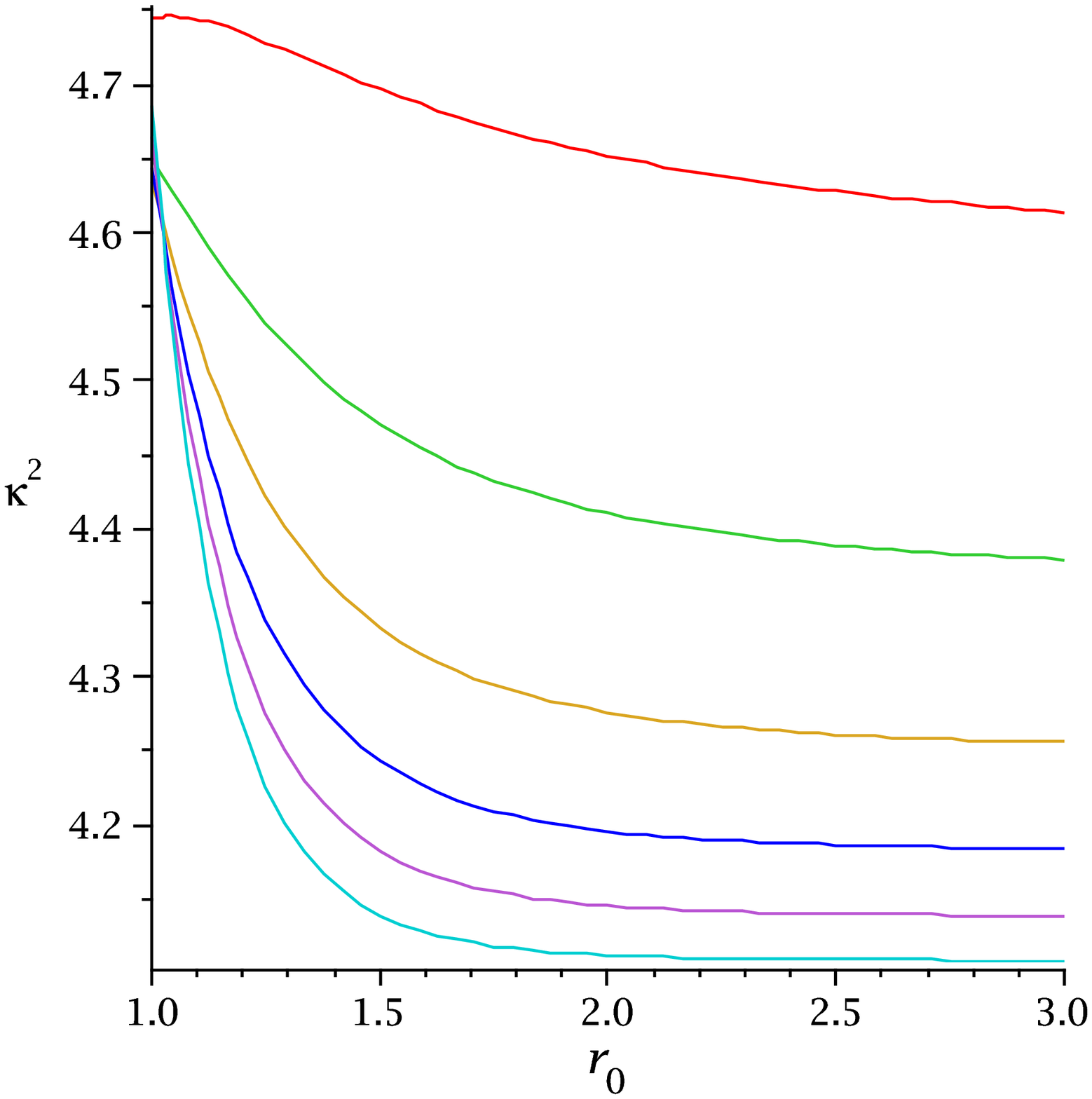}
 \caption{ Non-GR branch, $\Lambda \ne 0$ case.
   The frequencies  $\nu^2$  
   and $\kappa^2$  for $\Lambda=1$, $C=1$,
   $\tilde{\alpha}=1,$ $1 \le r_0 \le 3$, and $n= 5,$(top curve) $6, 7, 8, 9, 10$ (bottom curve). In this case we have stable orbits of arbitrary size.} \label{fig8to9}
 \end{figure}

 Figure (\ref{fig8to9}) shows  the functions
  $\nu^2$  and $\kappa^2$  for $\Lambda=1$, $C=1$,
  $\tilde{\alpha}= 1,$ $1 \le r_0 \le 3$, and $n= 5,$(top curve) $6, 7, 8, 9, 10$ (bottom curve). In this case we have stable orbits of arbitrary size.

We find a similar behavior  for the equivalent case with $\Lambda=-1.$

\begin{table}
\centering
\begin{tabular}{|c|c|c|c|}
\hline \hspace{2mm} Newtonian case \hspace{2mm} & \hspace{2mm} $n=4$
\hspace{2mm} & \hspace{2mm} $n=5$ \hspace{2mm} & \hspace{2mm} $n \ge
6$ \hspace{2mm} \\ \hline \hspace{2mm} $\Lambda = 0$ \hspace{2mm} &
\hspace{2mm} SSC \hspace{2mm} & \hspace{2mm} NSC
\hspace{2mm} & \hspace{2mm} NSC \hspace{2mm} \\
 \hspace{2mm} $\Lambda < 0$ \hspace{2mm} & \hspace{2mm} SSC \hspace{2mm} & \hspace{2mm} SSC\hspace{2mm} & \hspace{2mm} SSC \hspace{2mm} \\
 \hspace{2mm} $\Lambda > 0$ \hspace{2mm} & \hspace{2mm} SSC \hspace{2mm} & \hspace{2mm} NSC\hspace{2mm} & \hspace{2mm} NSC \hspace{2mm} \\
\hline
\end{tabular}
\begin{minipage}[t]{12cm}
\caption{Stability of circular orbits in $n$-dimensional Newtonian
  gravity with or without cosmological constant. The existence of some stable curves is denoted by SSC, and their not existence by NSC.}
  \lb{table1}
\end{minipage}
\end{table}

\

\begin{table}
\centering
\begin{tabular}{|c|c|c|}
\hline   \hspace{4mm} GR case  \hspace{4mm} & \hspace{4mm} $n=4$     & \hspace{4mm}  $n \ge 5$ \hspace{4mm}\\ 
\hline \hspace{4mm} $\Lambda = 0$ \hspace{4mm} & \hspace{4mm} SSC \hspace{4mm} & \hspace{4mm}  NSC \hspace{4mm} \\
       \hspace{4mm} $\Lambda < 0$ \hspace{4mm} & \hspace{4mm} SSC \hspace{4mm} & \hspace{4mm}  SSC \hspace{4mm} \\
       \hspace{4mm} $\Lambda > 0$ \hspace{4mm} & \hspace{4mm} SSC \hspace{4mm} & \hspace{4mm} NSC\hspace{4mm} \\
\hline
\end{tabular}
\begin{minipage}[t]{12cm}
\caption{Stability of circular orbits in n-dimensional Einstein
  gravity with or without cosmological constant. The existence of some stable curves is denoted by SSC, and their not existence by NSC.}  \lb{table2}
\end{minipage}
\end{table}

\begin{table}
\centering
\begin{tabular}{|c|c|c|c|c|c|c|}
\hline  EGB  & \multicolumn{2}{c|}{$\Lambda = 0$} &
 \multicolumn{2}{c|}{$\Lambda > 0$} & \multicolumn{2}{c|}{$\Lambda <
 0$} \\  \cline{2-7} ($\alpha \ne 0$) & $\alpha>0$ & $\alpha<0$ & $\alpha>0$ &
 $\alpha<0$ & $\alpha>0$ & $\alpha<0$ \\ \hline non-GR branch &
 SSC& NSC& SSC & NSC & SSC & NSC\\ \hline GR branch & SSC & NSC & SSC
 &  NSC & SSC & SSC \\ \hline
\end{tabular}
\caption{Stability of circular orbits in Einstein-Gauss-Bonnet theory
  with or without cosmological constant and different signs for the
  EGB coupling constant. The existence of some stable curves is denoted by SSC, and their not existence by NSC.
} \lb{table3}
\end{table}

\section{Summary and Final Remarks}

{ An observationally viable theory of gravitation must allow for
  the existence of stable limited orbits. It is known that, in the
Newtonian gravity as well as in  classical general relativity, these
orbits exist only if the dimension of the usual space is three
($n=4$). We analyze the linear stability of circular geodesics in a
$n$-dimensional ($n \ge 4$) spacetime with not null cosmological
constant $\Lambda$ and we find that there exist cases in which these
orbits are stable. 

We also analyse the linear stability of circular geodesics for
$n$-dimensional spherically symmetric static solutions of
Einstein-Gauss-Bonnet (EGB) gravity. Here we also find that there
exist cases with stable circular orbits. The differential equation
that gives us the metric function has two families of solutions
corresponding to the sign in front of the square root in
(\ref{fsolution}). Following the definition given at \cite{nozawa},
the family with minus (plus) sign is called GR branch (non-GR branch)
solution. The EGB part of our work is divided in several cases,
depending on whether the cosmological constant $\Lambda$ is zero or
not, and on the kind of solution used (GR or non-GR branch solution).

For the non-GR solution, the sign of the coupling parameter $\alpha$
is very important. When $\alpha <0$, no matter the value of $\Lambda$
or $n$, it is not possible to find stable circular orbits. Otherwise,
when $\alpha >0$, for all $n \ge 5$, it is possible to find stable
circular orbits, no matter whether $\Lambda$ is null, positive or
negative.

The results obtained for the GR solution are similar. When $\Lambda$
is positive or null, and $\alpha$ is negative, the circular orbits are
not stable. But we show values of $\alpha>0$, $5\le n \le 9$, $\Lambda
\ge 0$ and $r_0$ such that the existence of stable circular orbits is
possible.  The great difference of this branch solution occurs in the
case $\Lambda<0$. In this case, for example, taking $\Lambda=-1$,
$\alpha=\pm 1$, $n\ge 5$ and $r_0 \ge 1.2$ we have stability.  The
results above summarized in tables (\ref{table1}-\ref{table3}).

 A similar  analysis as the one presented in this article 
is under study for  higher dimensions rotating  
black holes like the one presented in Refs. \cite{aliev} (five-dimensional GR with magnetic field),
\cite{page} (n-dimensional GR) and \cite{radu} (five-dimensional EGB theory).
Another possible generalizations and applications of the present work concerns with the study of gravitational
radiation in higher dimensions in the case where circular orbits do
exist. For gravitational radiation in $n$-dimensional spacetimes, see for instance\cite{lemos}. Also the  ultrarelativistic plunge into
 a multidimensional black hole was study in \cite{emanuele}.
A generalization of this calculation to circular orbits
may be relevant to improve our understanding of mini-black hole production in
TeV-gravity scenarios.

\acknowledgements

V.M.R. thanks the hospitality of DMA-IMECC-UNICAMP. P.S.L. thanks the
partial financial support of FAPESP and CNPq. Also we thank Dr. J. Schleicher for his help with the English translation of the Kant's text.


\end{document}